\documentclass[12pt,onecolumn]{IEEEtran}
\renewcommand{\baselinestretch}{1.0}
\usepackage[colorlinks=false]{hyperref}
\usepackage[dvips]{graphicx}
\usepackage[usenames,dvipsnames]{color}
\usepackage{epsfig}
\usepackage{rotating}
\usepackage{amssymb}
\usepackage{amsmath,amsfonts}
\usepackage[boxed]{algorithm}
\usepackage[latin1]{inputenc}
\usepackage{verbatim,comment}
\usepackage[tableposition=top,font=small,labelfont=bf,format=hang]{caption}
\usepackage{booktabs}

\newtheorem{thm}{Theorem}
\newtheorem{prop}{Proposition}

\newtheorem{cor}{Corollary}

\newcommand{\w}{ \textsf{\it w}}

\newcommand{\N}{{\mathbb{N}}} 

\newcommand{\R}{{\mathbb{R}}} 
\newcommand{\C}{{\mathbb{C}}}

\newcommand{\CC}{\mathcal{C}}

\newcommand{\II}{\mathcal{I}}

\newcommand{\Aut}{\operatorname{Aut}}

\newcommand{\Trace}{\operatorname{Trace}}

\DeclareMathOperator{\wt}{wt}

\begin{document}

\title{On Bounded Weight Codes}
\author{Christine Bachoc  \hspace{.2cm}Venkat Chandar \hspace{.2cm}G\'erard
Cohen \hspace{.2cm}  Patrick Sol\'{e} \hspace{.2cm}  Aslan
Tchamkerten\thanks{C. Bachoc is with the University of Bordeaux (France), V.
Chandar is with MIT Lincoln Laboratory (USA), and  G.~Cohen , P.~Sol\'e, and
and A.~Tchamkerten are with Telecom ParisTech (France).  A. Tchamkerten is partly supported by an Excellence
Chair grant from the French National Research Agency (ANR, ACE project). Part of
this work appeared at ISIT 2010.}}
\maketitle
\begin{abstract}
The maximum size of a  binary code is studied as a function of its length $n$,
minimum distance $d$, and {\emph{minimum}} codeword weight $ \w$.
This function $B(n,d,\w)$ is first characterized in terms of its exponential growth rate in the limit $n\rightarrow
\infty$ for fixed $\delta=d/n$ and
$\omega=\w/n$. The exponential growth rate of $B(n,d,\w)$ is shown to be
equal to the exponential growth rate of $A(n,d)$ for $0\le \omega \le 1/2$, and equal to the
exponential growth rate of $A(n,d,\w)$ for $1/2<\omega\le 1$.  Second, analytic and numerical
upper bounds on $B(n,d,\w)$ are derived using the semidefinite programming (SDP)
method. These bounds yield a non-asymptotic improvement of the second Johnson bound and  are tight for certain values of the parameters.
\end{abstract}
{\keywords Constant weight codes, Johnson bounds, semidefinite programming}
\section{Introduction}

Two classical functions in combinatorial coding theory are $A(n,d)$, the largest size of a
binary code of length $n$ and minimum distance $d$, and $A(n,d,\w)$, the largest size of a
binary code of length $n$, minimum distance $d$, and constant weight $\w$. A closely
related function is $B(n,d,\w)$, obtained from $A(n,d,\w)$ by relaxing the weight
constraint to only require that the weight of each codeword is at least $\w$. Codes
satisfying a minimum weight constraint are called {\it heavy weight codes} in
\cite{CST10},
where they are motivated by certain asynchronous communication problems. The other
relaxation where codewords are required to have weight at most $\w$ defines the function
$L(n,d,\w)$. Complementation immediately shows that $L(n,d,\w)=B(n,d,n-\w)$. The function $L$
naturally occurs in the proof of the Elias bound \cite[Lemma 2.5.1]{HP}. It also occurs in
the problem of list decoding when bounding the size of the list as a
function of the decoding radius $\w$. In this problem, $L(n,d,\w)$ represents the largest size of a list of codewords
at distance at most $\w$ from the received vector,  given a binary code of length $n$ and minimum
distance $d.$ This function is denoted by $A'_2(n,d,\w)$ in \cite{GS}, where the Elias Lemma \cite[Lemma
2.5.1]{HP} is referred to as the Johnson bound, and is used to prove upper bounds on the list size.

In the present paper we first characterize the asymptotic exponent of
$B(n,d,\w)$ as a function of those of $A(n,d)$ and $A(n,d,\w)$ (Theorem~\ref{oo}). This result is
based on the  asymptotic unimodality of $A(n,d,\w)$, which was conjectured in
\cite[Conjecture $2$]{CST10}. Note that, the non asymptotic analogue of this
result  (posed as a research problem in \cite[p.674]{MS}) is {\em false} as
$A(15,6,6) < A(15,6,7)$ \cite{O}.

Second, we provide upper bounds on $L(n,d,\w)$ obtained by the semidefinite
programming method. From these bounds, we derive a non asymptotic improvement
of the Elias/Johnson Lemma in a certain range of $n$, $d$, and $\w$ (Theorem~\ref{Bound degree 2}) as well as
numerical tables.

The material is organized as follows. Section~\ref{prelim} contains elementary
bounds and some tables of $B(n,d,\w)$ derived therefrom. Section~\ref{results}
contains the asymptotic results. Section~\ref{sdepe} is dedicated to the SDP
method. Section~\ref{constructions} explores three heavy weight codes
construction techniques. In Section~\ref{conclusion} we provide some concluding
remarks.
\section{Elementary bounds}
\label{prelim}
In this section we establish a few basic relations between $B(n,d,\w)$ and
$A(n,d,\w)$. 

Note first that $B(n,d,\w)$ is increasing in
$n$, and decreasing in $d$ and $\w$. Further,
by definition of $B(n,d,\w)$, we have 
\begin{equation}\label{basic3}
B(n,d,\w)\ge A(n,d,j)\quad \text{for  } j\ge \w\,.
\end{equation}

By taking weight classes sufficiently far
apart so that they do not overlap, we get
\begin{equation}\label{basic4}
B(n,d,\w)\ge \sum_{h=0}^{\lfloor
\frac{n-\w}{d}\rfloor} A(n,d,\w+hd)
\end{equation}
where $\left\lfloor x\right\rfloor$ denotes the largest integer not exceeding~$x$.

Since any code is a disjoint union of
 constant weight codes, we have
\begin{equation}\label{basic1}
B(n,d,\w)\le \sum_{j=\w}^n A(n,d,j)\,.
\end{equation}

Removing the weight constraint can only improve the size, hence
\begin{equation}\label{basic2}
B(n,d,\w)\le A(n,d)=B(n,d,0)\,.
\end{equation}

The following result is analogous to the first half of the first Johnson bound
\cite[($3a$)]{sloane2}:\footnote{Whether or not the analogous of the second half of the first Johnson bound,
i.e. \cite[($3b$)]{sloane2}, holds as well remains an open question. Specifically,
it is unclear at this point whether the inequality
$$B(n,d,\w)\le \frac{n}{n-\w}
B(n-1,d,\w)$$
is valid. }
{\prop \label{John} For $\w\le n$ we have $$B(n,d,\w)\le \frac{n}{\w}
B(n-1,d,\w-1)\,.$$
 }
\begin{IEEEproof}
Let $C$ be a code realizing $B(n,d,\w)$, and consider the matrix
whose rows are its codewords. Since the average weight of a column, which we
denote by $W$, is given by
the total number of $1$'s in the matrix divided by $n$, we get
\begin{align}\label{eq:exist}
W\geq \frac{\w B(n,d,\w)}{n}\,.
\end{align}
Now, say column $l$ has weight at least $W$ (one such column clearly exists).
Pick the subcode of
$C$ given by the codewords of $C$ that have a $1$ in the $l$-th position.
Modify this subcode by deleting the $l$-th component of each
codeword. If we denote by $C'$ the resulting code,
we conclude that $W\le |C'|\le B(n-1,d,\w-1).$ Using this together with
\eqref{eq:exist} yields the desired result.
\end{IEEEproof}

Finally, the following Gilbert type lower bound is immediate:
{\prop  \label{pr1} For all $n\geq 1$, $d\leq n$, and $\w\leq n$ $$B(n,d,\w )\ge \frac{\sum_{i=\w}^n
{n \choose i}}{\sum_{i=0}^{d-1}{n \choose
i}}\,.$$  }

We conclude this section with tables derived from the preceding bounds.
Some trivial entries are $B(n,d,\w)=1$ whenever $d>\min\{2\w,2(n-\w)\}$.  We
limited $n$ and $d$ to the values where $A(n,d)$ and $A(n,d,\w)$ are known exactly
(for all $\w$) in  \cite{sloane1,sloane2}. 
Entries of the tables where $\w>n$ are left blank.

\begin{table}[h]
\centering
\tiny{
\centering
\caption{$B(n,4,\w)$}
\label{d=4}
\begin{large}
\begin{tabular}{|c|c|c|c|c|c|c|c|c|c|} \hline
$n$ & $A(n,4)$  &$\w=2$ &$\w=3$&$\w=4$ &$\w=5$  &$\w=6$  &$\w=7$  &$\w=8$  &$\w=9$  \\ \hline
$6$ &   4            &        4    &         3-4   &        3 -4   &           1  &           1  &              &             &   \\
$7$  &   8            &        8 &      7-8   &        7-8  &        3-5   &          1   &            1  &             &    \\
$8$ &      16          &          16  &      15 -16         & 15-16  &         8-10   &     4 -6       &    1         &          1    &           \\
$9$  &       20         &         20   &        19-20    &       19-20     &        18-20    &    12-18       &       4-6       &         1  &        \\
 $10$ &       40        &      40      &      39- 40     &    39-40        &        36-40     &        30-40     &        13-20      &    5-7       &      1       \\ \hline
\end{tabular}
\end{large}
}
\end{table}
\begin{table}[h]
\centering
\tiny{
\centering
\caption{$B(n,6,\w)$}
\label{d=6}
\begin{large}
\begin{tabular}{|c|c|c|c|c|c|c|c|c|c|} \hline
$n$ & $A(n,6)$  &$\w=2$ &$\w=3$&$\w=4$ &$\w=5$  &$\w=6$  &$\w=7$  &$\w=8$  &$\w=9$  \\ \hline
$9$  &      4          &      4     &      4      &      3- 4      &       3- 4    &        3-4     &       1     &1             &  1      \\
 $10$ &      6         &       6     &         6   &       6     &       6      &          5- 6  &      3- 6       &   1         &   1     \\
 $11$ &      12          &       12     &    12       &    11-12        &      11-12       &   11- 12         &    6- 9         &      3-  6     &  1         \\
  $12$&      24          &       24       &      24        &     23- 24        &      23-24         &   23-24            &   12-  24      &     9-16        &    4- 7        \\
  $13$&       32           &        32     &      32       &       31-32      &     31- 32        &      31- 32       &       26-32       &       18-32      &  13- 20        \\ \hline
\end{tabular}
\end{large}
}
\end{table}
\begin{table}[h]
\centering
\tiny{
\centering
\caption{$B(n,8,\w)$}
\label{d=8}
\begin{large}
\begin{tabular}{|c|c|c|c|c|c|c|c|c|c|} \hline
$n$ & $A(n,8)$  &$\w=2$ &$\w=3$&$\w=4$ &$\w=5$  &$\w=6$  &$\w=7$  &$\w=8$  &$\w=9$  \\ \hline
  $12$&        4        &   4         &      4      &      4      &       4      &        4     &    3- 4         &  3- 4          &     1       \\
  $13$&          4        &  4          &     4       &   4         &      4       &        4     &     4         &       3-4      &       3-4    \\
$14$  &          8       &    8       &      8      &    8        &       8      &       8      &     8         &   7- 8         &  4- 8          \\
$15$  &          16         &       16       &      16       &     16        &       15-16        &      15-16         &     15-16           &   15-16           &         10- 16      \\ \hline
\end{tabular}
\end{large}
}
\end{table}

\section{Asymptotics}\label{results}
For fixed $\delta,\omega\in [0,1]$, we denote by $b(\delta,
\omega)$ the exponential growth rate of  $B(n,d,\w)$ with respect to $n$ with $d=d(n)=\left\lfloor \delta n \right\rfloor$ and
$\w =\w(n)=\left\lfloor \omega n \right\rfloor$, i.e.
$$b(\delta,\omega)=\limsup_{n\rightarrow \infty}\left(\frac{1}{n}\log
B(n,d(n),\w (n))\right)$$
where logarithms are taken to the base $2$ throughout the paper. The asymptotic exponents of $A(n,d,\w)$
and $A(n,d)$ are defined similarly and are denoted by $a(\delta, \omega)$ and
$a(\delta)$, respectively. 


{\prop  For any $\delta\in [0,1]$ and
$\omega \in [0,1/2]$, we have $b(\delta,
\omega)=a(\delta).$  }
\begin{IEEEproof}
The Elias-Bassalygo bound \cite[equation ($2.8$)]{MRRW}
\begin{align}\label{ebass}
\frac{A(n,d)}{2^n}\le \frac{A(n,d,\w)}{{n\choose \w}}
\end{align}
together with the trivial inequality
$ A(n,d,\w)\le A(n,d)$
shows that the asymptotic exponents of $A(n,d)$ and $A(n,d,n/2)$ are the same. 
The result then follows by combining the bounds 
(\ref{basic3}) and (\ref{basic2}) to obtain
$$A(n,d,n/2) \le B(n,d,\w)\le A(n,d)$$
for $\w\le n/2$.
\end{IEEEproof}

The next result provides the main ingredient for proving that 
$b(\delta,\omega)=a(\delta,\omega)$ when $\omega \in (1/2,1]$.
{\thm \label{thbas} For fixed $\delta\in [0,1]$, $a(\delta,\omega)$  is 
unimodal in $\omega$ with a maximum at $\omega=1/2$.}
{\cor \label{oo}For any $\delta \in [0,1]$ and  $\omega \in (1/2,1]$, we have
$b(\delta,\omega)=a(\delta,\omega).$}
\begin{IEEEproof}[Proof of Corollary~\ref{oo}]
 We have 
 \begin{align}\label{igor}\max_{ j\in
\{\w,\w+1,\ldots,n\}} A(n,d,j) \leq B(n,d,\w)\le (n-\w+1) \max_{ j\in
\{\w,\w+1,\ldots,n\}} A(n,d,j) 
\end{align}
 by \eqref{basic3} for the first inequality and
by \eqref{basic1} for the second inequality. Letting $\w=\lfloor\omega n\rfloor$
and $d=\lfloor\delta n\rfloor$
we get
\begin{align}
\max_{ j\in
\{\w,\w+1,\ldots,n\}} A(n,d,j)=\max_{\rho\in [\omega,1]}A(n,d,\lfloor \rho
n\rfloor)\,,
 \end{align}
and therefore from~\eqref{igor} we have
$$ b(\delta,\omega)= \sup_{\omega\le \rho \le 1}a(\delta,\rho) $$
for any $\delta\in [0,1]$ and
$\omega \in [0,1]$. Assuming that $1/2< \omega\leq 1$, the theorem then follows
from Theorem \ref{thbas}.
 \end{IEEEproof}
\begin{IEEEproof}[Proof of Theorem~\ref{thbas}:]
We establish that $a(\delta,\cdot)$ is non-decreasing  over $ 
[0,1/2]$. This, by complementation, shows that  $a(\delta,\cdot)$ is
non-increasing over  $[1/2,1]$, proving the claim.

 Fix $\delta\in [0,1]$ and let  $\omega_1,\,\omega_2$ be such that $0\leq
\omega_1 <\omega_2\leq 1/2$. Throughout the proof we disregard discrepancies
due to the rounding of non-integer quantities as they play no role
asymptotically. Thus, for instance, we shall always treat $\omega_1n$ as 
if it is an integer. 

We show that,
from a given constant weight code $C_1$  with parameters $(n,d=\delta
n,\w_1=\omega_1n)$  such that $|C_1|=A(n,d,\w_1)$, it is possible
to construct a constant weight code $C_2$  with parameters
$(n,d,\w_2=\omega_2n)$, of size at least equal to $|C_1|$ multiplied 
by $1/(n+1)^2$. This shows that $a(\delta,\omega_2)\ge
a(\delta,\omega_1)$. The code $C_2$ is obtained from $C_1$ via translation. 

For a given fixed codeword  ${\mathbf{c}} \in C_1$, let us construct
a length $n$ binary vector ${\mathbf{t}}$ of weight 
$$\w=\omega n =\frac{\omega_2-\omega_1}{1-2\omega_1}n$$
 as follows. Consider first the
positions of ${\mathbf{t}}$ that form the support of ${\mathbf{c}}$ ($\w_1$ of
them). Pick 
$\omega_1\w$ of these positions arbitrarily and assign them $1$'s. Similarly, assign $1$'s
to an arbitrary selection of the
$(1-\omega_1)\w$ positions that lie outside the support of
${\mathbf{c}}$. The remaining positions of 
${\mathbf{t}}$ are filled with $0$'s. Note that, by of our choice of $\w$, the vector ${\mathbf{c}}'={\mathbf{t}} \oplus
{\mathbf{c}}$ (component
wise modulo $2$ sum of ${\mathbf{t}} $ and  ${\mathbf{c}}$) has weight
$\w_2$.

Now observe that, because the selections made to construct 
${\mathbf{t}}$ are arbitrary, for any given ${\mathbf{c}} \in C_1$ there are 
$${\omega_1n \choose \omega_1\omega n}{(1-\omega_1)n \choose (1-\omega_1)\omega
n}$$
ways of choosing ${\mathbf{t}}$ for which ${\mathbf{c}}'$
has weight $\w_2$. Therefore, if we now pick ${\mathbf{t}}$ randomly and
uniformly among
all possible sequences of weight $\w$, the probability that this sequence
translates a given ${\mathbf{c}} \in C_1$ to a sequence of weight $\w_2$ is given by
$$p=\frac{{\omega_1n \choose \omega_1\omega n}{(1-\omega_1)n \choose
(1-\omega_1)\omega n}}{ {n \choose \omega n}}\,.$$
This implies that a vector ${\mathbf{t}}$ that is randomly and
uniformly chosen among
all possible sequences of weight $\w$ translates on average
$$p A(n,d,\w_1)$$
codewords from  $C_1$ into codewords of weight $\w_2$ (and minimum distance
$d$). Therefore, 
$$A(n,d,\w_2)\geq p A(n,d,\w_1) \,.$$
Finally, using the following standard bounds on binomial
coefficients\footnote{$h(p)$ denotes the binary entropy $-p\log p-(1-p)\log
(1-p)$.} 
$$\frac{1}{(n+1)} 2^{n h(k/n)}\leq {n \choose k}\leq 2^{n h(k/n)}\quad k\leq
n\,,$$
(see, e.g.,\cite[Example 11.1.3,
p.353]{CT}) shows that $$p\geq \frac{1}{(n+1)^2}\,.$$
Therefore we obtain $$A(n,d,\w_2)\geq \frac{1}{(n+1)^2} A(n,d,\w_1)\,,$$
from which the theorem follows.\end{IEEEproof}

\section{Upper bounds on $L(n,d,\w)$ from semidefinite programming}
\label{sdepe}


The semidefinite programming  method is a far reaching
generalization of Delsarte linear programming method to obtain bounds
for extremal problems in coding theory. In the present situation, we
aim at upper bounding $L(n,d,\w)$, which is the maximal number of
elements of a code contained in the ball $B(\w)$ centered at the
all-zero word with radius $\w$ of the binary Hamming space
$H_n=\{0,1\}^n$. We obtain numerical bounds for small values of the
parameters $(n,d,\w)$, which improve the elementary bounds for
$B(n,d,n-\w)=L(n,d,\w)$ given in Section II. We also obtain a new bound,
which is an explicit function of $(n,d,\w)$, and improves on the
Elias/Johnson bound for some values of these parameters. 

The numerical bounds are obtained by a straightforward application of
the SDP method. We refer to \cite{ITW} for a survey of this method and its applications to the binary
Hamming space, including the case of codes in balls. See also
\cite{BGSV} for a survey on the more general subject of symmetry
reduction of semidefinite programs, with applications to coding
theory. In a few words,
$L(n,d,\w)$ can be interpreted as the independence number of a certain graph with vertex
set $H_n$, thus is upper bounded by the so-called {\em Lov\'asz theta number} $\vartheta$ of this graph
(or rather  by its strengthening $\vartheta'$),
which is the optimal value of a certain semidefinite program. This SDP
has exponential size, but can be reduced to polynomial size by the
action of the symmetry group of the graph, which is the symmetry group
of $B(\w)$, i.e. the group $S_n$ of permutations of the
$n$ coordinates. 

Let us recall that a function $F:H_n^2\mapsto \R$ is said to be {\em positive
definite} (or positive semidefinite)  if the matrix $(F(x,y))$ indexed
by $H_n$ is positive semidefinite.  This property is denoted $F\succeq
0$. In the symmetrization process discussed above, a description of
the $S_n$-invariant positive definite functions on $H_n$ is required. 
This description is in fact provided in \cite{S}, under the
name of block diagonalization of the Terwilliger algebra of the
Hamming space, and in the framework of group representations in
\cite{V}.
Numerical upper bounds for $L(n,d,\w)$ obtained in this
way are displayed in Tables IV, V, VI.

For the announced explicit bound, we use a slightly different (and
self contained)
formulation of the SDP bound, which is given in Theorem \ref{Theorem}. We
shall recover the Elias/Johnson bound as a special case, and obtain a
new bound in Theorem \ref{Bound degree 2}.
There, we follow  the same line for Hamming balls as the one followed
for spherical caps in \cite{Bachoc-Vallentin}. In the latter, the SDP
method has lead to numerical bounds and also to explicit bounds of
degree
up to two.

\subsection{Improving the Johnson bound}

We start with a more handy restatement of the SDP bound, which is
essentially the dual form of the SDP defining the theta number $\vartheta'$. 
The notations are as follows: the space of functions on $H_n$ is denoted 
$\CC(H_n)=\{f:H_n\mapsto \C\}$ and is endowed with the standard 
inner product $\langle f_1,f_2\rangle=\frac{1}{2^n}\sum_{x\in H_n}
f_1(x)\overline{f_2(x)}$. We shall consider the decomposition of this
space
under the action of the full automorphism group $\Aut(H_n)$ of the
Hamming space and under the action of the symmetric group $S_n$. Since the irreducible components
are indeed real, we can restrict to the real valued functions.

The orbit  of  $(x,y)\in H_n^2$ under the action of $S_n$ is
determined uniquely by  the values
of  $u:=\wt(x)$, $v:=\wt(y)$ and $t:=d(x,y)$. Thus the elements of
$F\in \CC(H_n^2)$ which are $S_n$-invariant,  i.e. which satisfy $F(gx,gy)=F(x,y)$ for all $g\in S_n$,
$(x,y)\in H_n^2$, are of the form $F=F(u,v,t)$.
With this notation, $F\succeq 0$ stands
for:
$(x,y)\mapsto F(\wt(x),\wt(y),d(x,y))\succeq 0$. 

{\thm\label{Theorem}
Let
\begin{equation*}
\Omega(n,d,\w):=\{ (u,v,t)\in \N^3 : \ 0\leq u,v\leq \w, \ d\leq t\leq n,$$ $$
t\leq u+v, \ u+v-t\equiv 0\mod 2\}.
\end{equation*}
Let $P(u,v,t)\in \R[u,v,t]$ be a polynomial symmetric in $(u,v)$.
If $P$ satisfies the following conditions:
\begin{enumerate}
\item $P-f_0\succeq 0$ for some $f_0>0$
\item  $P(u,v,t)\leq 0$ for all $(u,v,t)\in \Omega(n,d,\w)$,
\item $P(u,u,0)\leq 1$ for all $u\in \{0,\dots,\w\}$,
\end{enumerate}
then
\begin{equation*}
L(n,d,\w)\leq \frac{1}{f_0}.
\end{equation*}
} 
\begin{IEEEproof}
For $(x,y)\in H_n^2$, let $F(x,y):=P(\wt(x),\wt(y),d(x,y))$. We consider for
a code $C\subset B(\w)$ with minimal distance at least equal to $d$,  the sum
\begin{equation*}
S:=\sum_{(x,y)\in C^2} F(x,y).
\end{equation*}
From property (1) of $P$, we have $S\geq f_0|C|^2$. On the
other hand, $S=S_1+S_2$ where $S_1$ is the sum over pairs $(x,y)\in
C^2$ with $x=y$ and $S_2$ is the sum over the non equal pairs
$(x,y)\in C^2$, $x\neq y$. Condition 2) on $P$ insures that $S_2\leq 0$ and
condition 3) on $P$ that $S_1\leq |C|$. Altogether we obtain $|C|\leq 1/f_0$.
\end{IEEEproof}

In order to apply the above theorem with specific polynomials
$P(u,v,t)$,
we need an explicit  description of those who are positive definite.
 Such a description is indeed obtained in
\cite{S}, and in \cite{V}  in terms of orthogonal polynomials (Hahn polynomials to
be precise). As we shall see, for our purpose, we need a slightly
different expression. 

A general method is explained in \cite{B}, \cite{ITW}, \cite{BGSV},
involving group representation. The space $\CC(H_n)$ can be decomposed
into the direct sum of $S_n$-irreducible subspaces. The sum of those
subspaces which are isomorphic to a given irreducible representation
of $S_n$ is called an isotypic subspace. 
We recall that certain matrices $E_k(x,y)$
are associated to the isotypic components $\II_k$ of $\CC(H_n)$ under
the action of $S_n$. Here $k\in [0..\lfloor n/2\rfloor]$, 
$\II_k$ corresponds to the irreducible
representation $[n-k,k]$ of the symmetric group $S_n$, and has
multiplicity
$n-2k+1$. Moreover, $E_k(x,y)$ is $S_n$-invariant thus can be
expressed in terms of $(u,v,t)$, namely $E_k(x,y):=Y_k(u,v,t)$.
Then we have the following characterization (we use the standard
notation $\langle A,B\rangle=\Trace(AB^*)$ for matrices):
{\prop
For all $P\in \R[u,v,t]$, symmetric in $(u,v)$, $P\succeq 0$ if and
only if
\begin{equation}
P(u,v,t)=\sum_{k=0}^{\lfloor n/2\rfloor} \langle F_k ,E_k(x,y)\rangle
\end{equation}
where for $k\in [0..\lfloor n/2\rfloor]$, $F_k\in \R^{m_k\times m_k}$, $m_k=n-2k+1$, and $F_k\succeq 0$.
}
More precisely,
$E_k(x,y)$ is computed 
from a decomposition
of $\II_k$ into irreducible subspaces
$\II_k=R_{k,1}\oplus \dots R_{k,m_k}$. If for all $i$, $(e_{k,i,1},\dots
,e_{k,i,h_k})$ is an orthonormal basis of $R_{k,i}$ in which the action
of $S_n$ is expressed by the same matrices (i.e., not depending on $i$), then
\begin{equation*}
E_{k,i,j}(x,y)=\sum_{s=1}^{h_k} e_{k,i,s}(x)e_{k,j,s}(y).
\end{equation*}
The decomposition of $\II_k$  with irreducible submodules is not
unique but changes $E_k(x,y)$ to
$AE_k(x,y)A^*$ for an invertible matrix $A$, see \cite[Lemma
  4.2]{B}. Note that such a change does not affect the above
characterization
of $P$ being positive definite since
$\langle F_k, AE_k(x,y)A^*\rangle =\langle A^*F_kA, E_k(x,y)\rangle$
and $F_k\succeq 0$ if and only if $A^*F_kA\succeq 0$. 

There are essentially two
strategies to obtain such a decomposition. One can start from the
decomposition of $X=H_n$ into orbits under the action of $S_n$,
namely $X=X_0\cup\dots \cup X_n$, with $X_k=\{x\in H_n : \wt(x)=k\}$,
which leads to a decomposition of the functional space
$\CC(X)=\CC(X_0)\perp\dots\perp \CC(X_n)$ and then decompose each
$S_n$-space $\CC(X_k)$, following \cite{D}. It is the method adopted in
\cite{V} where the corresponding matrices $E_k(x,y)$ are obtained in
terms of Hahn polynomials. Another approach starts from the
decomposition of $\CC(H_n)$ under the full $\Aut(H_n)$, namely
$\CC(H_n)=P_0\perp P_1\perp\dots\perp P_n$ where $P_k=\oplus_{\wt(\w)=k}
\C \chi_\w$, $\chi_\w(x)=(-1)^{\w\cdot x}$, then decomposes each $P_k$ under the action of the subgroup
$S_n$.
Because we want to work with polynomials in $(u,v,t)$ of low degree,
this last decomposition is better suited. Indeed, if $P\in \R[u,v,t]$,
then
$x\mapsto F(x,y):=P(\wt(x),\wt(y),d(x,y))$ belongs to $P_0\perp\dots\perp
P_k$ if and only if the total degree of $P$ in the variables $(u,t)$
is at most equal to $k$. 

An isomorphism of $S_n$-modules between $\CC(X_k)$ and $P_k$ is given
by $\phi_k$:
\begin{align*}
\phi_k: \CC(X_k) &\to P_k\\
   f &\mapsto \phi_k(f) :=\sum_{\wt(\w)=k} f(\w)\chi_\w.
\end{align*}
so we have exactly the same picture for the decomposition of
$\CC(H_n)$ when $P_k$ replaces $\CC(X_k)$, namely
the irreducible decomposition of $P_k$ under the
action of $S_n$ that is  for  $0\leq k\leq \lfloor \frac{n}{2}\rfloor$, we have
\begin{equation}\label{dec1}
P_k=H_{0,k}\perp H_{1,k}\perp\dots \perp H_{k,k}
\end{equation}
and the isotypic components of $\CC(H_n)$, i.e. 
\begin{equation*}
\II_k=H_{k,k}\perp H_{k,k+1}\perp\dots\perp H_{k,n-k}\simeq
H_{k,k}^{n-2k+1}.
\end{equation*}
Since $u=\wt(x)$, as a function of $x$, is invariant under $S_n$, and
is of degree $1$, the isotypic subspace $\II_k$ can also be decomposed
as:
\begin{equation*}
\II_k= \oplus_{i=0}^{n-2k} u^i H_{k,k}
\end{equation*}
Moreover, starting from an orthonormal  basis $(e_{k,s})$ of $H_{k,k}$, we obtain
an orthonormal basis $( u^i e_{k,s})$ of $u^i H_{k,k}$ in which the action of $S_n$ is
expressed by the same matrices, thus we can use it to compute the
corresponding matrix $E_k(x,y)$ the coefficients of which will be
equal to:
\begin{equation*}
E_{k,i,j}(x,y)=u^iv^j \sum_{s=1}^{h_k} e_{k,s}(x)e_{k,s}(y).
\end{equation*}
In other words, it is enough to compute $Z_k(x,y):=\sum_{s=1}^{h_k}
e_{k,s}(x)e_{k,s}(y)$, which is the zonal function associated to $H_{k,k}$, in terms of $(u,v,t)$. We obtain:

{\prop
We have the following expressions for $Z_k$, up to a positive
multiplicative constant:
\begin{itemize}
\item $Z_0=1$
\item $Z_1=-t+u+v-2uv/n$
\item $Z_2=t^2
  +(2/(n-2))(n-nu-nv+2uv)t+(1/(n-1)(n-2))(4u^2v^2-4n(u^2v+uv^2)+(n+2)(n-1)(u^2+v^2)+2n(n+1)uv-2n(n-1)(u+v))$
\end{itemize}
}

\begin{IEEEproof} We take the following notations:
if $\wt(\w)=1$, and $\w_i=1$, we let $\chi_i:=\chi_{\w}$. Let
\begin{equation*}
\begin{cases}
U:=n-2u=\sum_{i=1}^n \chi_i(x),\\
V:=n-2v=\sum_{i=1}^n \chi_i(y),\\
T:=n-2t=\sum_{i=1}^n \chi_i(x)\chi_i(y).
\end{cases}
\end{equation*}
Following \cite{D}, and the isomorphism $\phi_k$ defined above, 
$H_{k,k}=\ker(d)$ where $d: P_k\to P_{k-1}$ is defined by:
$d\chi_\w=\sum \chi_{\w'}$ where the sum is over the words $\w'$ of
weight $\wt(\w')=\wt(\w)-1$, and of support contained in the support of $w$.
We set $d=d_x$ to specify the variable under consideration and 
$d=d_x+d_y$ when applied to a function $F(x,y)$ on $H_n^2$. Then,
$Z_k$ is uniquely determined up to a multiplicative constant by the
properties:
\begin{enumerate}
\item $Z_k\in \R[U,V,T]$, is symmetric in $(U,V)$,
\item $x\mapsto Z_k(x,y)$ belongs to $P_k$,
\item $dZ_k=0$.
\end{enumerate}
According to the decomposition \eqref{dec1} with pairwise non
isomorphic irreducible subspaces, the space of functions satisfying conditions (1) and (2)
below is of dimension $1+k$. In the  variable $x$, $U$ and $T$ belong
to $P_1$, and it is easy to check that $U^2-n$, $UT-V$, $T^2-n$, belong to $P_2$. Thus a basis
for the space of functions satisfying  (1) and (2) is given by:
\begin{equation*}
\begin{cases}
k=0: \quad\{1\}\\
k=1: \quad\{UV,T\}\\
k=2: \quad\{(U^2-n)(V^2-n),\\ UVT-U^2-V^2+n, T^2-n\}
\end{cases}
\end{equation*}
The assertion $Z_0=1$ is then trivial. In order to compute $Z_1$ and $Z_2$,
we need formulas for the image under $d$ of the monomials in
$(U,V,T)$. We compute the following:
\begin{equation*}
\begin{cases}
d_x 1=d1=0,\\
d_x U=n \quad \text{thus}\quad d(UV)= n(U+V),\\
d_x T= V \quad \text{thus}\quad d T=U+V.
\end{cases}
\end{equation*}
With the above we obtain that $Z_1$ is proportional to
$T-\frac{1}{n}UV$.
Similarly we obtain:
\begin{equation*}
\begin{cases}
d(U^2+V^2)=2(n-1)(U+V),\\
d(U^2V^2)=2(n-1)(U^2V+UV^2),\\
d(UVT)=(U^2V+UV^2)+(n-2)(U+V)T,\\
d(T^2)=-2(U+V)+2(U+V)T.
\end{cases}
\end{equation*}
and $Z_2$ turns to be proportional to
$$
T^2-n -\frac{2}{n-2} (UVT-U^2-V^2+n)$$ $$ +\frac{1}{(n-1)(n-2)} (U^2-n)(V^2-n).
$$
From the identity $Z_k(x,x)=\sum e_{k,s}(x)^2$, we have that
$Z_k(U,U,0)\geq 0$ which determines the sign of the multiplicative
factor.
We obtain the announced formulas.
\end{IEEEproof}

\noindent{{\it{Remark:}}} The method used to calculate the polynomials $Z_k$ for
  $0\leq k\leq 2$ outlines an algorithmic way to compute $Z_k$ for
  general $k$. It would be more satisfactory to have an expression of
these polynomials in terms of orthogonal polynomials.

Now we apply Theorem \ref{Theorem} in order to obtain upper bounds for
$L(n,d,\w)$.
We start with a polynomial $P(u,v,t)$ of degree one and recover Elias
bound:
Let 
\begin{align*}
P(u,v,t):=&Z_1(u,v,t)+d-2\w(1-\w/n)\\
=& d-t+(u+v-2uv/n) -2\w(1-\w/n).
\end{align*}
With $f_0:=d-2\w(1-\w/n)$, we have $P-f_0\succeq 0$. 
If $\w\leq n/2$, the maximum over $[0,\w]^2$ of $u+v-2uv/n$ equals
$2\w(1-\w/n)$, and is attained for $u=v=\w$. Thus $P(u,v,t)\leq 0$ for
$(u,v,t)\in \Omega(n,d,\w)$,
and $P(u,u,0)\leq d$. Thus we obtain that {if }$\w\leq n/2$ { and } $d> 2\w(1-\w/n),$ then
\begin{equation}\label{Elias}
  L(n,d,\w)\leq
\frac{d}{d-2\w(1-\w/n)}.
\end{equation}

It is unclear in general how to design a good polynomial $P$ of degree
$k$. A
possible strategy is to start from a  polynomial $L(t)$ optimizing the
bound for $A(n,d)$ and disturb it with a polynomial $p(u,v)$, i.e.
take $P=L(t)+p(u,v)$. Since $L(t)\succeq 0$, 
condition (1) of Theorem \ref{Theorem}, is equivalent to $F_0-f_0E_0\succeq 0$. 
In order to fulfill condition (2), it is enough to have $p(u,v)\leq 0$
for $[u,v]\in [0,\w]^2$ so one can take $p(u,v)=(u+v-2\w)s(u,v)$ or
$p(u,v)=(u(u-\w)+v(v-\w))s(u,v)$ where $s(u,v)$ is a sum of squares. 
For the degree $1$, if one follows this line and takes
$P=(d-t)+\lambda(u+v-2\w)$ with $\lambda>0$, one finds that the optimal
choice of $\lambda$ is $\lambda=1-2\w/n$ and obtains again the Elias
bound \eqref{Elias}. For the degree $2$, we consider accordingly a
polynomial $P$ of the form
\begin{equation*}
P=(t-d)(t-n) +\lambda (u(u-\w)+v(v-\w)),
\end{equation*}
with $\lambda\geq 0$.
The matrix $F_0(\lambda)$ associated to $P$ is equal to
\begin{equation*}
F_0(\lambda)=\begin{pmatrix}
nd & -n-d-\lambda \w & 1+\lambda \\
     & 4n/(n-1) +2d/n  & -4/(n-1)\\
        && 4/(n(n-1))\\
\end{pmatrix}.
\end{equation*}
Let $f_0(\lambda):=\det(F_0(\lambda)$. The lower left $2\times 2$ corner of $F_0(\lambda)$ is positive
semidefinite so the matrix $F_0(\lambda)-f_0E_0$ is
positive semidefinite if and only if its determinant is non negative,
which amounts to the condition 
\begin{equation*}
f_0\leq \frac{n^2(n-1)}{8d} f_0(\lambda).
\end{equation*}
On the other hand $$P(u,u,0)=dn+2\lambda u(u-\w)\leq dn$$ so we obtain
the bound $8d^2/((n-1)f_0(\lambda))$. It remains to find the maximum
of $f_0(\lambda)$, which is a polynomial of degree $2$
in $\lambda$:
$$
\frac{n(n-1)}{2} f_0(\lambda)= -((n-1)d+2(n-\w)^2)\lambda^2$$ 
$$+d(2n+2-4\w)\lambda +d(2d-(n-1)).
$$
The maximum is attained for $\lambda_0= d(n+1-2\w)/((n-1)d+2(n-\w)^2)$,
$\lambda_0\geq 0$ if $\w\leq (n+1)/2$, and is equal to
\begin{equation*}
\frac{4d\big(d^2+\frac{2(n-\w)(n+1-2\w)}{n-1}d-(n-\w)^2\big)}{n((n-1)d+2(n-\w)^2)}.
\end{equation*}
This last value is positive if and only if 
\begin{equation*}
d>\frac{(n-\w)}{(n-1)}\big(\sqrt{2(n-\w)(n-1)}-(n+1-\w)\big).
\end{equation*}
Altogether we obtain:

{\thm\label{Bound degree 2}
Assume $\w\leq (n+1)/2 $ and 
\begin{equation*}
d>\frac{(n-\w)}{(n-1)}\big(\sqrt{2(n-\w)(n-1)}-(n+1-\w)\big).
\end{equation*}
Then
\begin{equation*}
L(n,d,\w)\leq \frac{2d\big(d+\frac{2(n-\w)^2}{n-1}\big)}{d^2+\frac{2(n-\w)(n+1-2\w)}{n-1}d-(n-\w)^2}.
\end{equation*}}

\noindent{{\it{Example:}}} with  the above we obtain $L(n,n/2,n/2)\leq 2n-1$. It is an almost sharp bound in view of 
 $A(n,n/2,n/2)=2n-2$ for values of $n$ for which an Hadamard matrix of order $n$ exists \cite[Theorem 10]{sloane2}.
Note that adding the all zero codeword to such an Hadamard code yields $L(n,n/2,n/2)= 2n-1$.

\noindent{{\it{Example:}}} For $d=2\w(1-\w/n)$ the degree $1$ bound does not apply. The
degree $2$ gives a bound if $\w> n/2-\sqrt{n^2/(2(n+1))}$ which equals
\begin{equation*}
\frac{2\w(n^2-\w)}{\frac{n^2}{2}-(n+1)\big(\w-\frac{n}{2}\big)^2}.
\end{equation*}

\subsection{Tables}

The  tables IV, V and VI  give  upper bounds  of $L(n,d,\w)$ employing
the SDP method. They   {\emph{always}} improve on the bound (\ref{basic2}) (Cf right most column) and sometimes on (\ref{basic1}) when the latter is stronger than the former.
This situation is indicated by a star exponent.

 In some cases they allow us to derive {\emph{exact}} values of $L(n,d,\w)$ by using the expurgation technique of the 
next section. These cases are indicated by bold face numbers. To do that we collect the weight enumerators of some special binary codes in the notation of \cite{MS}.
\begin{table}\caption{$d=4$}
\begin{equation*}
\begin{array}{|c|c|c|c|c|c|c|c|c|c|c|c|c|}
\hline
n \backslash \w &4&5&6&7&8&9&10&11&12&13&& A(n,4) \leq \\
\hline
10 &31 &37 & & & & & & & & && 40 \\
11 &{42}^* &67 & & & & & & & & && 72\\
12 &{56}^* & 100 & 138 & & & & & & & && 144\\
13 & {72 }^*& {144}^* & 221 & 248 & & & & & & && 256\\
14 & {92 }^* &{201 }^* &340 & 411&486 &503 & & & & && 512\\
15 & {114 }^*& {274}^*& 508 & 750 & 849 & 989 & 1002& & & && 1024\\
16 & {\bf 141}^* & {365}^* & 736 & 1184& 1571& 1767 & 1984& 2012&& && 2048\\
17 & 171 &477 & 1039 & 1813 & 2602& 2981& & & & && 3276\\
18 & 205 & 613 & 1437 & 2703 & 4183 & 5041 & 6007 & 6324 & & && 6552\\
19 & 243 & 776 & 1947 & 3933 & 6541 & 9174 & 10532 & 12249 & 12641 & && 13104\\
20 & 286 & 970 & 2594 & 5600 & 9976 & 14966 & 19390 & 21965 &
24834 & 25388&& 26168\\
\hline
\end{array}
\end{equation*}
\end{table}

The weight enumerator of the $RM(2,4)$ dual of the $RM(1,4)$ is computed by MacWilliams transform \cite[Ch. 5, Th. 1]{MS} as
$$x^{16}+y^{16}+140(x^{12}y^4+x^4y^{12})+448(x^{10}y^6+x^6y^{10})$$ $$+870 x^8y^8.$$
This shows by expurgation that
$$L(16,4,4)=141.$$
\begin{table}\caption{$d=6$}
\begin{equation*}
\begin{array}{|c|c|c|c|c|c|c|c|c|c|c|c|c|c|c|c|}
\hline
n \backslash \w  &6&7&8&9&10&11&& A(n,6)\leq\\
\hline
14& 51 & 56 & 63 & &  &  &&64\\
15& 74 & 96 & 113 & 127 & & &&128\\
16& {\bf  113}& 157&207 &228 & {\bf 255}& 255 &&256\\
17& 159 & 250 & 318 & & & &&340\\
18& 205& 409& 481& 563& 677 &  &&680\\
19& 259& 554 & 752 & 913 & 1107 & &&1280 \\
20& 324& 739 & 1200 &1519 & 1835& 2096 &&2372\\
\hline
\end{array}
\end{equation*}
\end{table}
The weight enumerator of the Nordstrom Robinson code is
$$x^{16}+y^{16}+112(x^{10}y^6+x^6y^{10})+30 x^8y^8.$$
This shows by expurgation
$$L(16,6,6)=113,\,L(16,6,10)= 255 .$$
\begin{table}\caption{$d=8$}
\begin{equation*}
\begin{array}{|c|c|c|c|c|c|c|c|c|c|c|c|}
\hline
n \backslash \w  &8&9&10&11&12 & 13 & 14&15&16&&  A(n,8)\leq\\
\hline
18 & 67 & & & & & & &&&& 72 \\
19 &100 & 123 & 137& & & & &&&& 142 \\
20 & 154& 222 & 253& & & & &&&& 256\\ 
21 & 245 & 359 & 465  &   & & & &&&& 512\\ 
22 & 349 & 598 & 759 & 870 & 967 & 990 & 1023 &&&& 1024\\ 
23 & {\bf 507} & 831 & 1112 & 1541 & 1800 & 1843 & 1936 & 2047 &{\bf 2048} &&2048 \\ 
24 & {\bf 760} & 1161 & 1641& 2419& {\bf 3336} & 3439 & 3711 & 3933 & {\bf 4095} && 4096\\ 
\hline
\end{array}
\end{equation*}
\end{table}
The weight enumerator of the extended Golay code is 
$$x^{24}+y^{24}+759(x^{16}y^8+x^8y^{16})+2576 x^{12}y^{12}.$$
Shortening we obtained the dual of the perfect Golay code.
$$x^{23}+506x^{15}y^8+1288 x^{11}y^{12}+253 x^7 y^{16}.$$
This shows by expurgation
$$ L(24,8,8)=760,\, L(24,8,12)=3336,\,L(24,8,16)=4095,$$
and
$$ L(23,8,8)=507,\,L(23,8,16)=2048.$$

\section{Constructions}\label{constructions}
Three well studied code construction techniques are expurgation, translation, and  concatenation. In the context of heavy weight codes,
the first is perhaps mostly of theoretical interest as a good decoding algorithm needs
not, in general, provide a good decoding algorithm for a subcode.  In contrast,
the other two techniques also provide practical decoding algorithms.

\subsection{Expurgation}

The following result shows that, for $\w\leq d$, 
$B(n,d,\w)$ and $A(n,d)$ are essentially the same
(recall that $B(n,d,\w)\leq A(n,d)$).
{\prop \label{star}  For $1\leq \w\leq d \leq n$,  we have
$$B(n,d,\w)\ge A(n,d)-1.$$}
\begin{IEEEproof}
Let $C$ be a code achieving $A(n,d).$ By first translating this code so that to include
the all-zero codeword, then by removing the all-zero codeword, we get a new code
of size $A(n,d)-1$, with minimum distance  and
weight both at least equal to $d$. The proposition
follows.
\end{IEEEproof}
{\thm For all large enough and even $n$, all $\w\le
n/2$, and all $d  \le n
h^{-1}(1/2)$,\footnote{$h^{-1}(\cdot)$ denotes the inverse function of the
binary entropy over the range $[0,1/2]$.} we have $$ B(n, d, \w)\ge 2^{(n-2)/2}.$$}
\begin{IEEEproof}
Pick a self dual code above the Gilbert bound \cite{MST}. This code being binary
self-dual, contains the all-one codeword, and is
therefore self-complementary. Hence,  half of its
codewords at least have weight at least $ n/2.$
\end{IEEEproof}


\subsection{Translation}

 We assume that the reader has some familiarity with the covering radius
concept \cite{C+}. Recall that the covering radius of a code is the smallest
integer $t$ such that Hamming balls of radius $t$ centered on the codewords
cover the ambient space. Define $R(n,d)$ as the largest covering radius of a
code achieving $A(n,d).$ Since the covering radius exceeds
$\lfloor(d-1)/2\rfloor$, we get $R(n,d) \ge \lfloor(d-1)/2\rfloor$ with
equality iff the code that achieves $R(n,d)$ is perfect. A sharper bound on $R(n,d)$ for non perfect
codes is obtained as a direct consequence of the sphere covering
bound$$2^n\le
A(n,d)\sum_{i =0}^{R(n,d)}{n \choose i}.$$

The motivation for taking ``largest'' rather than ``smallest'' in the definition
of $R(n,d)$ is to have the best upper bound on $\w$ in the next Proposition,
which sharpens, in certain cases, Proposition \ref{star}.

{\prop \label{cr} Fix two integers $n\geq 1$ and $d\geq 1$. If $\w \le R(n,d) $  then $$B(n,d,\w)= A(n,d).$$ }

\begin{IEEEproof}
Pick a  code $C$  realizing $A(n,d).$ There exists a translate of $C$ of weight $\w$ as
long as $\w$ is less than or equal to the covering radius of $C$. This gives $B(n,d,\w)\ge A(n,d)$. The reverse inequality is (\ref{basic2}).
\end{IEEEproof}

\subsection{Concatenation}
Consider an heavy weight code of length $n,$ size $q,$ minimum weight $\w$, and
distance $d$. If we concatenate this code with a code of length $N$, size $M$, and  minimum distance $D$ over $GF(q)$, we get a binary code of length
$Nn$, weight at least $ \w N$, size $M$ and  minimum distance $dD$. Hence, provided $B(n,d,\w)\ge q,$ we see that
$$B(Nn,dD,\w N)\ge A_q(N,D)\,.$$
where $A_q(N,D)$ denotes the largest size of a code of length $N$ and minimum distance $D,$ over $GF(q).$
Efficient decoding algorithms for concatenated codes can be found in \cite{Du}.
\section{Concluding remarks}\label{conclusion}
We investigated $B(n,d,\w)$, defined as the largest number of codewords of
weight at least $\w$ and minimum distance $d$.
 The asymptotic exponent of $B(n,d,\w)$ is reduced to those of $A(n,d)$ or
 $A(n,d,\w)$, depending on $\w$. For finite values of the parameters, we
 obtained bounds on $B(n,d,\w)$ partly using the SDP method. As future research,
 it might be possible to find 
new exact values of $B(n,d,\w)$ by special constructions. In this direction, one
possibility is to investigate $R(n,d)$ defined in Section~\ref{constructions}.
\section*{Acknowledgement}
We thank Navin Karshyap for pointing out \cite{O}.

\end{document}